# *Challenges and Opportunities of Machine Learning for Monitoring and Operational Data Analytics in Quantitative Codesign of Supercomputers*

Thomas Jakobsche
University of Basel
thomas.jakobsche@unibas.ch

Nicolas Lachiche
University of Strasbourg
nicolas.lachiche@unistra.fr

Florina M. Ciorba
University of Basel
florina.ciorba@unibas.ch

This work examines the challenges and opportunities of Machine Learning (ML) for Monitoring and Operational Data Analytics (MODA) in the context of Quantitative Codesign of Supercomputers (QCS). MODA is employed to gain insights into the behavior of current High Performance Computing (HPC) systems to improve system efficiency, performance, and reliability (e.g. through optimizing cooling infrastructure, job scheduling, and application parameter tuning [1]).

In this work, we take the position that QCS in general, and MODA in particular, require close exchange with the ML community to realize the full potential of data-driven analysis for the benefit of existing and future HPC systems. This exchange will facilitate identifying the appropriate ML methods to gain insights into current HPC systems and to go beyond expert-based knowledge and rules of thumb. The full potential of ML for QCS is not realized today due to the absence of a set of standard and best practices. To this end, we identify the following **challenges** related to current use of ML for QCS: (a) definition of appropriate operational data to be collected, (b) preparation of data for ML, (c) identification of appropriate ML methods, (d) explainability of ML models, (e) transferability of ML models, (f) FAIR data, privacy and sharing concerns, and (g) data-owners' and machine-learners' perspectives. To address the above challenges, we formulate **opportunities** to bring ML expertise into QCS and facilitate close collaboration: (1) invite ML experts to various thematic panel discussions, (2) review recent advancements in ML, and (3) establish an Open-Data Challenge. HPC administrators, users, and researchers working on improving data-driven operations and quantitative codesign will benefit from the deployment of appropriate ML methods. Advancing ML solutions can leverage the full potential of the vast amounts and types of data being collected on HPC systems. Most MODA solutions in production still involve a human in the loop [2] which prevents the full realization of the vision of autonomous computing systems [7]. ML can enable autonomous responses that are scalable beyond today's manual or ML-assisted capabilities for system optimization.

## 1. Challenges

We describe the challenges that need to be addressed to realize the full potential of ML-based data-driven analysis for HPC center-collected data and QCS.
(a) It is oftentimes challenging for HPC researchers to define what are appropriate data to collect, the individual that a ML model counts/learns on, and the population that such individuals belong to.
(b) Currently there is no single best way to filter and prepare center-collected data for ML. This is due to the growing data collection capabilities on HPC systems [5][6], resulting in terabytes of high-dimensional, often non-linear, time-series data for each component of the system.
(c) It is not clear which ML models are suitable for HPC center-collected data, holding back the development of appropriate MODA solutions. There are also no best practices on hyperparameter tuning, performance measuring, suitable training data, and validation in the context of HPC and QCS.
(d) Absence of explainable ML solutions and missing knowledge behind a specific ML-based decision raise concerns for system operators and prevent them from employing automatic ML-based actions.
(e) It is an open question whether ML models and insights are transferable outside of the systems they are

trained on and if they are useful for the design of next-generation supercomputers.
(f) HPC researchers face the issues of making data FAIR [8] and resolving privacy and sharing concerns when accessing, analyzing, and publishing data and code.
(g) Disjoint perspectives and approaches to data analysis: (g.1) Starting with the data, **data owners** ask questions such as: *"What analysis can we do and what problem can we solve?"* and (g.2) Starting with the learning problem, **machine learners** ask questions such as: *"What is the appropriate data and method of analysis?"*. These perspectives need to be reconciled as they play a significant role in the design and development of ML solutions for HPC center-collected data and for QCS.

The **risks** of using ML in QCS without addressing the above challenges are:
(i) Development of ML-based MODA solutions that only work on curated datasets. These solutions tend to not translate to production systems and real use-cases as their results are often not reproducible.
(ii) Reluctance to use ML for automated response in production, due to non-explainability.
(iii) Misleading insights from ML models that result in inappropriate response and/or design decisions.
(iv) Researchers with a data owner's perspective will only search for problems and ML methods that fit the data versus finding appropriate data and methods for given problems that ML can help solve.

## 2. Opportunities

We identify three opportunities to **bring ML expertise to QCS** and advance data-driven analysis. Each opportunity is associated with an **objective** and a potential technical approach is also outlined.
(1) Invite ML experts to directly interact with the QCS communities interested in data-driven analysis of HPC center-collected data via panel discussions held at related annual events such as QCS at SC, HPCMASPA at Cluster, MODA at ISC, PMACS at Euro-Par, and others. Discussion topics can target best practices of collecting and preparing HPC data for ML, identifying appropriate ML models to deploy, and success stories of using explainable ML to autonomously optimize system operations.
(2) Review recent advancements in ML to reveal appropriate strategies to handle high-dimensional, non-linear, time-series data with a focus on the challenges detailed above. The review can identify areas that are promising for HPC data analysis. Topics to look out for are concept drift, uncertainty quantification, autoencoders, variance thresholding, resampling, dynamic time warping, and time-aggregation. This is an opportunity for all active researchers in the ML and QCS communities.
(3) Initiate an "Open-Data Challenge" (ODC) to facilitate the development of ML for QCS. We envision an anomaly detection challenge based on representative datasets from different centers. The datasets can be generated through executions of representative (proxy-)applications and the performance anomalies produced by the HPC Performance Anomaly Suite [4]. The design of the challenge should include ML experts in order to correctly define the format of the challenge, the desired insights about ML for HPC, and how the challenge should be formulated to get these insights. This challenge is also a way to address the above mentioned concerns regarding data sharing and transferable ML models. ODC can spawn reproducible work and reveal whether different approaches and ML models are necessary to address similar use-cases on different HPC systems. ODC can also include a call to explore portable ML models across multiple systems [3].

## 3. Conclusion

This position paper is an initial effort and a call to the HPC and ML communities to connect and collaboratively build and deploy best and standard practices of using ML for QCS. The close exchange between the QCS and ML communities will reconcile the data owners' and machine learners' perspectives and lead to the development of appropriate ML-based solutions that will benefit and improve the operations of existing and the design of future HPC systems. A first immediate action is to invite ML experts to upcoming events in QCS and MODA. A subsequent immediate action is to organize, launch, and evaluate the ODC.